\documentclass{aa}
\usepackage{graphicx}
\usepackage[varg]{txfonts}
\usepackage{cases}
\usepackage{natbib}

\begin{document}

\title{Preflare very long-periodic pulsations  observed in H$\alpha$ emission before the onset of a solar flare}

\author{Dong~Li \inst{1,2,3}\thanks{Corresponding author}, Song~Feng\inst{4}, Wei~Su\inst{5}, and Yu~Huang\inst{1}}

\institute{Key Laboratory of Dark Matter and Space Astronomy, Purple Mountain Observatory, CAS, Nanjing 210033, PR China \email{lidong@pmo.ac.cn} \\
           \and State Key Laboratory of Space Weather, Chinese Academy of Sciences, Beijing 100190, PR China \\
           \and CAS Key Laboratory of Solar Activity, National Astronomical Observatories, Beijing 100101, PR China \\
           \and Yunnan Key Laboratory of Computer Technology Application, Faculty of Information Engineering and Automation, Kunming University of Science and Technology, Kunming 650500, PR China \\
           \and MOE Key Laboratory of Fundamental Physical Quantities Measurements, School of Physics, Huazhong University of Science and Technology, Wuhan 430074, PR China \\}
\date{Received; accepted}

\titlerunning{Preflare-VLPs observed in H$\alpha$ emission before onset of a solar flare}
\authorrunning{Dong Li et al.}

\abstract {Very long-periodic pulsations during preflare phases
(preflare-VLPs) have been detected in the full-disk solar soft X-Ray
(SXR) flux \citep[see][]{Tan16}. They may be regarded as precursors
to solar flares and may  help us better understand the trigger
mechanism of solar flares.} {In this letter, we report a
preflare-VLP event prior to the onset of an M1.1 circular-ribbon
flare on 2015 October 16. It was simultaneously observed in
H$\alpha$, SXR, and extreme ultraviolet (EUV) wavelengths.} {The SXR
fluxes in 1$-$8~{\AA} and 1$-$70~{\AA} were recorded by the
Geostationary Operational Environmental Satellite (GOES) and Extreme
Ultraviolet Variability Experiment (EVE), respectively; the light
curves in H$\alpha$ and EUV~211~{\AA} were integrated over a small
local region, which were measured by the 1m New Vacuum Solar
Telescope (NVST) and the Atmospheric Imaging Assembly (AIA),
respectively. The preflare-VLP is identified as the repeat and
quasi-periodic pulses in light curves during preflare phase. The
quasi-periodicity can be determined from the Fourier power spectrum
with Markov chain Monte Carlo (MCMC)-based Bayesian
\citep[e.g.,][]{Liang20}.} {Seven well-developed pulses are found
before the onset of an M1.1 circular-ribbon flare. They are firstly
seen in the local light curve in H$\alpha$ emission and then
discovered in full-disk SXR fluxes in GOES~1$-$8~{\AA} and
ESP~1$-$70~{\AA}, as well as the local light curve in AIA~211~{\AA}.
These well-developed pulses can be regarded as the preflare-VLP,
which might be modulated by LRC-circuit oscillation in the
current-carrying plasma loop. The quasi-period is estimated to be
$\sim$9.3~minutes.} {We present the first report of a preflare-VLP
event in the local H$\alpha$ line and EUV wavelength, which could be
considered a precursor of a solar flare. This finding should
therefore prove useful for the prediction of solar flares,
especially for powerful flares.}


\maketitle

\section{Introduction}
Solar flares represent the rapid and violent  process of releasing
magnetic free energy by reconnection, which is often characterized
by a complex magnetic field \citep[see][for a review]{Benz17}. The
coupling of a complex magnetic structure and plasma during a solar
flare usually causes a quasi-periodic phenomenon, which is referred
to as the quasi-periodic pulsation \citep[QPP, see][for recent
reviews]{Nakariakov19a,Kupriyanova20}. It is a common oscillatory
feature in the light curve of solar flare, which was first detected
in X-ray and microwave emission \citep[e.g.,][]{Parks69} and later
discovered in nearly all electromagnetic radiation, such as radio
\citep{Ning05,Karlicky20}, extreme-ultraviolet
\citep[EUV,][]{Shen19,Yuan19}, X-ray \citep{Ning14,Dennis17}, and
even $\gamma$-ray \citep{Nakariakov10,Li20a}. It could be observed
in the preflare phase \citep[e.g.,][]{Zhou16,Li20b}, rising and
postflare phases \citep{Kolotkov15,Li17,Ning17,Hayes19}. The
detected periods in solar flares vary from sub-seconds to hundreds
of seconds, which strongly depend on the observed instruments and
wavelengths \citep[e.g.,][]{Tan07,Shen13,Inglis16,Li16,Pugh19,Yu19}.
The QPPs can be found in most flare events, however, their
generation mechanisms are still highly debated \citep[see][for
reviews]{Van16,McLaughlin18}, which might be attributed to
magnetohydrodynamic (MHD) waves
\citep{Anfinogentov15,Wang15,Tian16,Nakariakov19b} or repetitive
magnetic reconnection \citep{Kliem00,Thurgood17,Li20c}. Previous
observations also found that the periods could depend on the
mechanism producing them \citep[e.g.,][]{Tan10}. The short periods,
such as 10$-$100~s, detected in hard X-ray or $\gamma$-ray etc are
usually associated with the nonthermal electrons or ions that have
been accelerated by the repetitive reconnection
\citep[e.g.,][]{Aschwanden95,Nakariakov10,Li20a}. The long periods
(i.e., 5$-$40~minutes), however, are often attributed to the MHD
waves; for instance, the transverse oscillations observed in coronal
loops, which were often interpreted as the kink-mode waves
\citep[e.g.,][]{Nakariakov99,Duckenfield19}, as well as the standing
slow-mode waves detected in hot ($>$6~MK) loops measured by the
SUMER spectrometer, which were referred as SUMER oscillations
\citep{Wang02,Wang03,Wang11}.

In recent years, the studies of flare-related QPPs have achieved
significant progress in understanding the dynamic of solar flares,
so they must be taken account when constructing the flare model
\citep{Kupriyanova20}. Moreover, the QPP has been observed during
preflare phase, which could be regarded as the precursor of a solar
flare, in other words, as a convenient precursory indicator for the
powerful (M- or X-) flare \citep[e.g,][]{Tan16,Zhou16}. Therefore,
investigating QPPs before the onset of solar flares can help us to
understand their trigger mechanism and origin source. Using soft
X-ray (SXR) fluxes recorded by the Geostationary Operational
Environmental Satellite (GOES), \cite{Tan16} first reported the QPPs
with typical periods of 8$-$30~minutes during preflare phases and
they referred to them as very long pulsations in the preflare phase
(preflare-VLPs). On the other hand, QPPs in H$\alpha$ emissions have
also been reported, such as sausage oscillations in the cool
postflare loop \citep[e.g.,][]{Srivastava08}, and the multiple
periodic oscillations in newly formed loops following small-scale
magnetic reconnection \citep[e.g.,][]{Yang16}. The detection of QPPs
in H$\alpha$ emissions could provide essential information for
understanding MHD waves in the solar chromosphere \citep{Jess15}.
However, the preflare-VLPs in H$\alpha$ emission are rarely
reported. In this letter, we investigate a preflare-VLP event in
H$\alpha$, SXR and EUV wavebands. The preflare-VLP shares a same
source with the accompanied M1.1 flare, implying it could be the
precursor of the main flare. This letter is organized as follows:
Section~2 describes the observations and Section~3 presents our main
results. The conclusion and discussion are summarized in Section~4.

\section{Observations}
On October 16, 2015, an M1.1 circular-ribbon flare took place in the
active region (AR) NOAA 12434 (S11E45). The energy partition of this
circular-ribbon flare has been studied in detail by \cite{Zhang19}.
In this study, we focus on the preflare phase before the onset of
the circular-ribbon flare, that is, between
$\sim$05:01~UT$-$06:10~UT. It was simultaneously observed by GOES,
as well as the 1m New Vacuum Solar Telescope \citep[NVST,][]{Liu14},
Atmospheric Imaging Assembly \citep[AIA,][]{Lemen12}, and the
Extreme Ultraviolet Variability Experiment \citep[EVE,][]{Woods12}
onboard the Solar Dynamics Observatory (SDO).

The NVST is a one meter aperture vacuum telescope located at Fuxian
Solar Observatory, which is operated by Yunnan Observatory of the
Chinese Academy of Sciences. It mainly provides high resolution
images in H$\alpha$ and TiO channels \citep{Liu14}. In this study,
H$\alpha$ level1 data at the wavelength of 6562.8~{\AA} between
$\sim$05:01~UT and $\sim$06:10~UT were used to investigate the
preflare-VLP. They were processed by the frame selection (lucky
imaging) from a large number of short-exposure images
\citep[see][]{Tubbs04,Liu14,Xu14,Xiang16}. The H$\alpha$ images in
the line center are used here, which have a spatial scale of
$\sim$0.16~pixel$^{-1}$ and a time cadence of $\sim$48~s. We also
use the SXR fluxes recorded by GOES and the EUV SpectroPhotometer
(ESP) for SDO/EVE, and the EUV image at AIA~211~{\AA}.

\begin{figure}[]
\centering
\includegraphics[width=\linewidth,clip=]{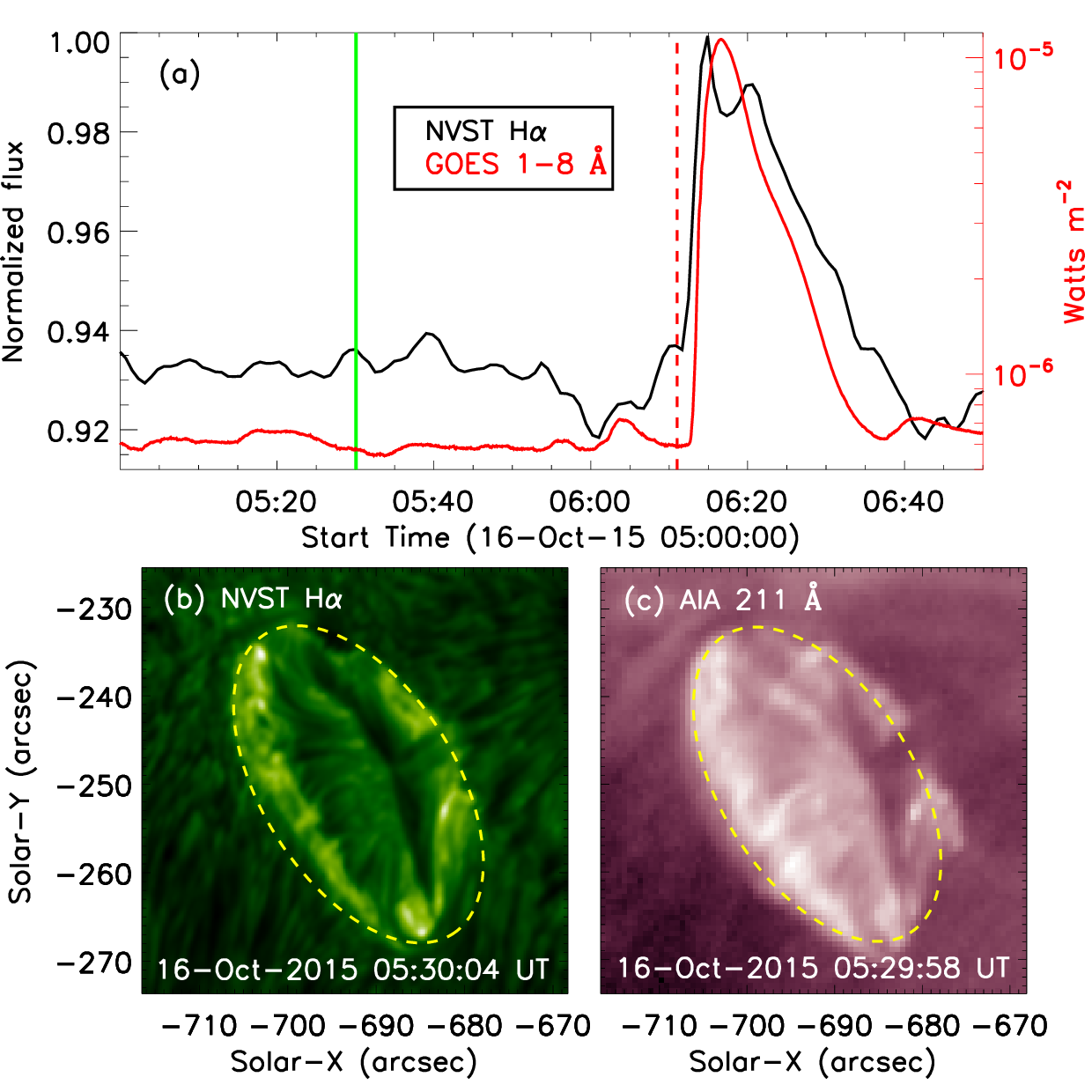}
\caption{ Panel~(a): Light curve integrated over a small local
region ($\sim$48\arcsec$\times$48\arcsec) in H$\alpha$ line center
(black) and the full-disk flux in SXR 1$-$8~{\AA} (red) from
05:00~UT to 06:50~UT on 2015 October 16. The red vertical line
indicates the onset of the M1.1 flare. Panels~(b) and (c): Snapshots
with a FOV of $\sim$48\arcsec$\times$48\arcsec at wavelengths of
H$\alpha$~6562.8~{\AA} and AIA~211~{\AA} at about 05:30~UT, as
indicated by the green line in panel~(a). The yellow dashed line
outlines an ellipse profile. The whole evolution is shown in the
accompanying video,  anim.mp4. \label{flux}}
\end{figure}

Figure~\ref{flux}~(a) presents the SXR flux in GOES 1$-$8~{\AA}
(red), which shows a swift enhancement at around 06:11~UT (red
vertical line), suggesting an M1.1 flare erupts. Some small pulses
before the onset of the M1.1 flare can be seen, which might be
regarded as the preflare-VLP \citep[e.g.,][]{Tan16}. We then plot
the light curve in H$\alpha$ emission integrated over a local region
with a small field-of-view (FOV) of about
48\arcsec$\times$48\arcsec, as shown with the black curve. We can
find a series of pronounced pulses in the preflare phase. Panels~(b)
and (c) draw the local images at around 05:30~UT (green vertical
line in panel~a) in H$\alpha$ and AIA~211~{\AA}, respectively. They
have the same FOV and display a circular profile, indicating a
followed circular-ribbon flare \citep[see,][]{Zhang19,Zhang20}. The
whole evolution from $\sim$05:01~UT to $\sim$06:20~UT can be seen in
the accompanying video, anim.mp4.

\section{Results}
\begin{figure}[]
\centering
\includegraphics[width=\linewidth,clip=]{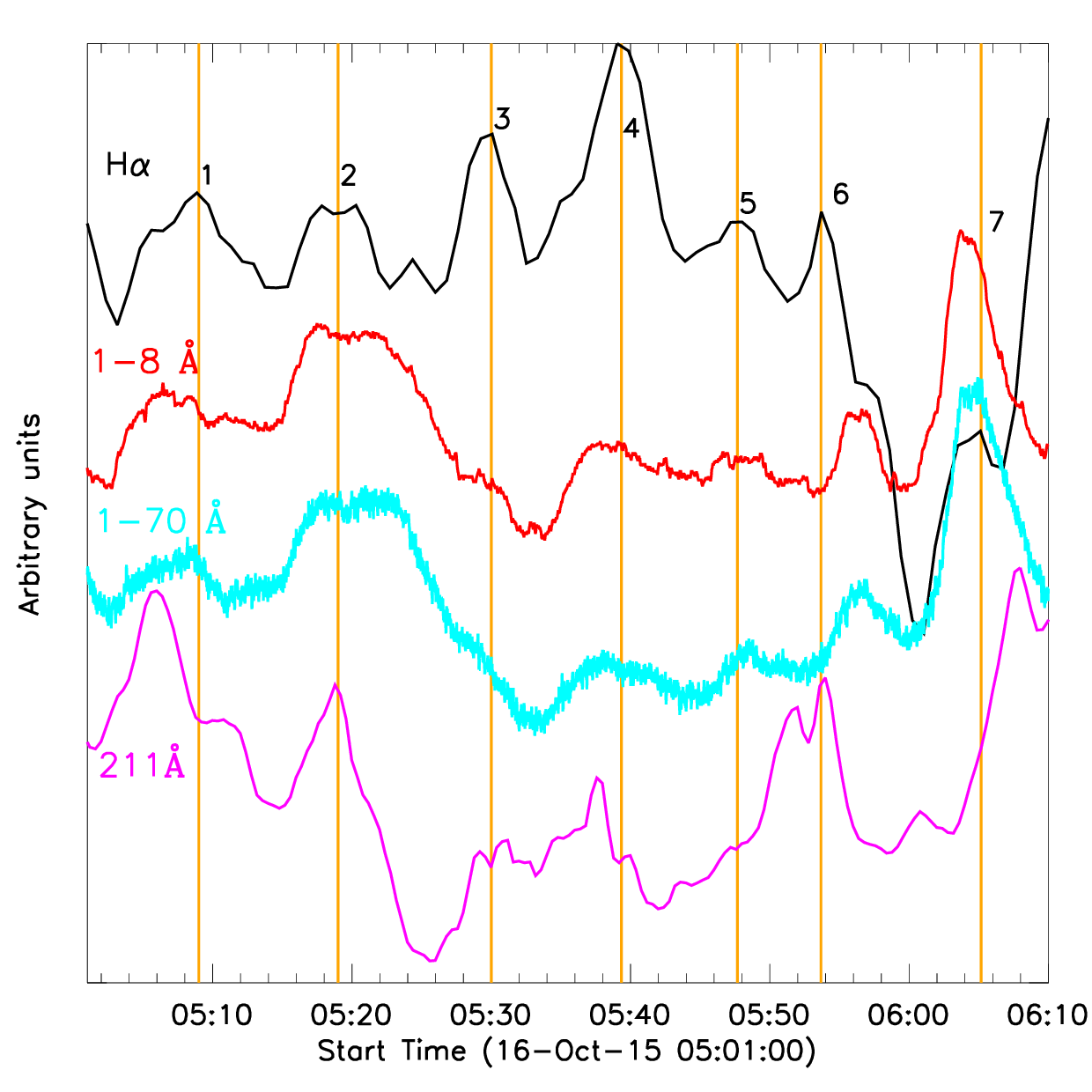}
\caption{Normalized light curves between 05:01~UT and 06:10~UT in
H$\alpha$~6562.8~{\AA} (black), GOES~1$-$8~{\AA} (red),
ESP~1$-$70~{\AA} (cyan), and AIA~211~{\AA} (magenta). The orange
lines mark the pulsation peak time in the H$\alpha$ light curve.
We note that the light curves have been shifted in the $y$~axis so they
can clearly be shown  in the same window. \label{curve}}
\end{figure}

Taking a closer look at the small pulses in the preflare phase,
Figure~\ref{curve} shows the local light curves during
05:01$-$06:10~UT in H$\alpha$ (black) and AIA~211~{\AA} (magenta),
as well as the full-disk light curves in GOES~1$-$8~{\AA} (red) and
ESP~1$-$70~{\AA} (cyan). All these light curves have been normalized
and shifted in height so that they may be clearly shown in the same
window. We can find seven well-developed pulses in the local
H$\alpha$ light curve, each of them assigned a number, and their
peak times are marked by orange vertical lines. The pulse period is
estimated to be in the range of 6$-$11.5~minutes, with an average
periodicity of $\sim$9.3~minuets. We also notice that the pulse
period from peak "1" to peak "5" is roughly constant, but it changes
clearly after peak "5." In the full-disk SXR fluxes at
GOES~1$-$8~{\AA} and ESP~1$-$70~{\AA}, we can find the seven pulses
with the period that is similar to the H$\alpha$ pulses. Moreover,
the two SXR fluxes are almost in phase with each other. We also note
that  pulse "3" is weak at GOES~1$-$8~{\AA} and it is hard to
distinguish in ESP~1$-$70~{\AA}. Then seven major pulses are
identified in the local light curve at the wavelength of
AIA~211~{\AA}, with the similar period for these pulses in the
H$\alpha$ light curve.

\begin{figure}[]
\centering
\includegraphics[width=\linewidth,clip=]{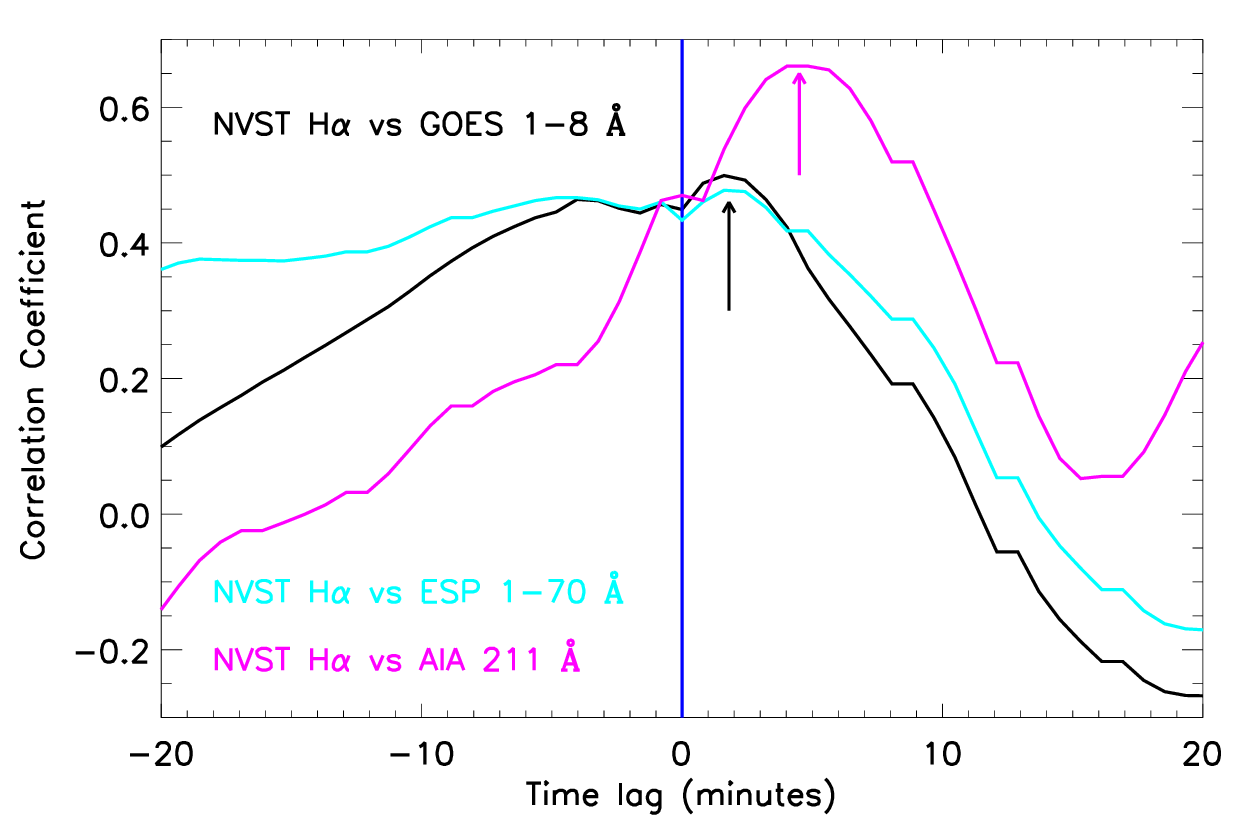}
\caption{Correlation coefficients between two light curves as a
function of the time lag, such as NVST H$\alpha$ and
GOES~1$-$8~{\AA} (black), ESP~1$-$70~{\AA} (cyan), and AIA~211~{\AA}
(magenta). A blue vertical line indicates at the time lag of "0".
\label{lag}}
\end{figure}

To determine the time delay in different wavelengths, the
cross-correlation analysis \citep[][etc.]{Tian12,Tian16,Su16} is
applied to these normalized light curves, as shown in
Figure~\ref{lag}. A maximum correlation coefficient of $\sim$0.5 can
be found at the time lag of roughly 1.8~minutes (marked by a black
arrow), suggesting a short time delay between the H$\alpha$ light
curve and the GOES~1$-$8~{\AA} flux. The same time delay can be
found between the H$\alpha$ light curve and the ESP~1$-$70~{\AA}
flux (cyan), which also implies that the pulses in SXR fluxes
recorded by GOES and EVE are fully in phase. We also find a maximum
correlation coefficient of $\sim$0.66 at the time lag of roughly
4.5~minutes (marked by a magenta arrow), indicating a long time
delay between the H$\alpha$ light curve and the AIA~211~{\AA} flux.

\begin{figure}[]
\centering
\includegraphics[width=\linewidth,clip=]{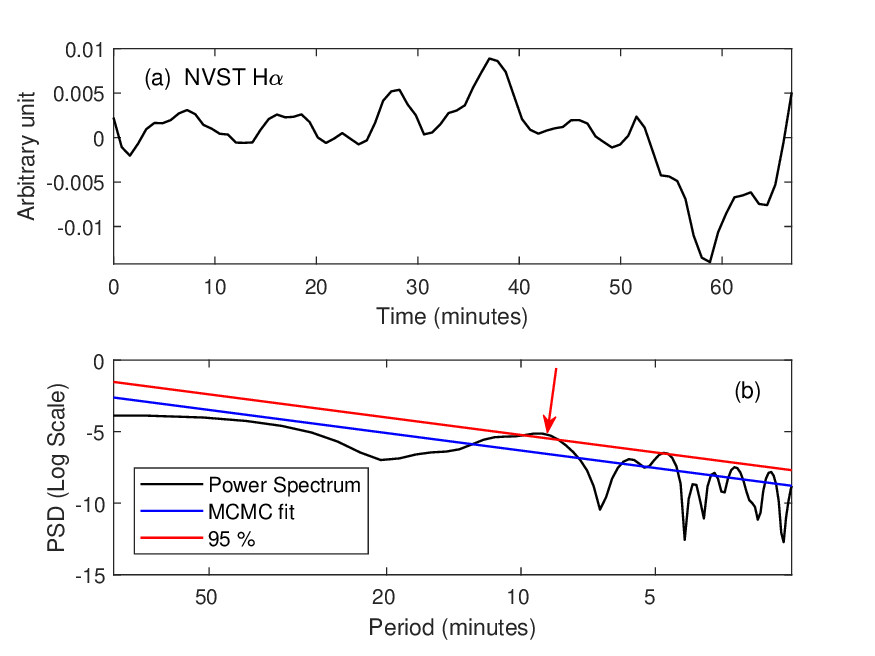}
\caption{Normalized H$\alpha$ light curve (a) and its corresponding
PSD (b) in log-log space. The blue line represents the best (MCMC)
fit, while the red line indicates the 95\% confidence level. The red
arrow marks the period which is above the confidence level.
\label{mcmc}}
\end{figure}

The pulse period is estimated by directly counting the pulses before
the onset time of the M1.1 flare (Figure~\ref{curve}), which is
straightforward and quite simple \citep[see,][]{Tan16}. To examine
their periodicity, we then perform a mathematical analytic method to
the original light curves, that is, a fast Fourier transformation,
as shown in Figure~\ref{mcmc}. The red noise here is estimated with
Multi-parameter Bayesian inferences based on MCMC samples
\citep[see,][]{Liang20}. Panel~(a) shows the H$\alpha$ light curve
during $\sim$05:01$-$06:10~UT and it has been normalized by
$(I-\overline{I})/\overline{I}$, where $I$ and $\overline{I}$ are
the observational intensities and their average intensity,
respectively. The power spectral density (PSD) of the normalized
light curve are given in panel~(b), displayed in a log-log space.
Based on this, a period of about 9.3~minutes is clearly found
 to be above the 95\% confidence level (red line), as
indicated by a red arrow. The MCMC result is consistent with
previous findings based on a direct count of the pulses.

\section{Conclusions and discussion}
Using the NVST data, a quasi-periodic pulsation event with a period
of $\sim$9.3~minutes was discovered in H$\alpha$ emission prior to
the onset of an M1.1 circular-ribbon flare on 2016 October 16. A
similar periodic pulsation is also detected in the full-disk SXR
fluxes recorded by the GOES and the SDO/EVE, as well as the local
EUV flux measured by the SDO/AIA. Based on the SXR emission observed
by the GOES, the very long periodic pulsations prior to the onset of
solar flares were reported by \cite{Tan16} and referred to as
preflare-VLPs. Here, the periodic pulsation event observed at the
wavelengths of H$\alpha$~6562.8~{\AA}, AIA~211~{\AA},
GOES~1$-$8~{\AA} and ESP~1$-$70~{\AA} in the preflare phase could
also be regarded as a preflare-VLP. Moreover, the H$\alpha$ and AIA
imaging observations suggest that the preflare-VLP shares a same
source origin region with the accompanying circular-ribbon flare,
further confirming that the preflare-VLP ought to be considered a
precursor of the main flare. Similar to the preflare coronal dimming
\citep{Zhang17} and the chromospheric evaporation in flare precursor
\citep{Li18}, the preflare-VLP cold be used to predict a solar
flare; in particular, it can be regarded as a precursory indicator
for the powerful flare \citep{Tan16,Zhou16}.

It is very interesting to note that a preflare-VLP has been
simultaneously observed in the H$\alpha$, SXR, and EUV wavebands.
The detected period is similar to the SUMER oscillations that
strongly damped \citep{Wang02,Wang03,Wang11}. However, the
preflare-VLP is not significantly damping, so it could not be
interpreted as the SUMER oscillation. On the other hand, the flare
source region might accumulate magnetic energy by photospheric
convection during the preflare phase, which can drive electric
currents in the plasma loop \citep{Tobias13,Tan06,Tan16}. Therefore,
the preflare-VLP is most likely to be explained as the LRC
oscillation in the current-carrying plasma loop, which can modulate
both thermal and nonthermal emissions
\citep[e.g.,][]{Tan10,Tan16,Li16,Li20b}. Previous studies
\citep{Zaitsev98,Zaitsev00} suggested that the LRC oscillation
period (P) could depend on the cross-sectional area (S), the plasma
density ($\rho$), and the electric current (I), such as
$P~\approx~(2.75~\times~10^{4}~S~\rho^{0.5})~/~I$
\citep[see,][]{Tan16}. The cross-sectional area could be estimated
from the source region in H$\alpha$ images, which is fitted with a
elliptic function, as outlined by the yellow dashed line in
Figure~\ref{flux}~(b) and (c). This is based on the fact that the
outer profiles of the source region hardly expand over time in the
chromosphere \citep[see][]{Zhang19}, which can also be seen in the
accompanying video, anim.mp4. The elliptic area following a
correction of  the projection effect is about
1.2$\times$10$^{14}$~m$^2$. The typical value in the flaring coronal
loop is referred to as the plasma density, that is,
$\rho\sim1.67~\times~10^{-11}$~kg~m$^{-3}$
\citep{Bray91,Tan16,Tian16}. Considering a period of
$\sim$9.3~minutes, the electric current in the preflare phase is
estimated to be roughly $I~\approx~2.4~\times~10^{10}$~A. The
estimated electric current here is lower than those taking place
during solar flares, that is, as high as $\sim$10$^{12}$~A
\cite[e.g.,][]{Canfield93,Tan06}. However, we should state that the
preflare-VLP appears prior to the onset of a solar flare, which also
agrees with previous findings in preflare phases
\citep[see,][]{Tan16}, so this result is reasonable.

In this letter, we first report a preflare-VLP in the H$\alpha$
emission, which could be regarded as a precursor to an M1.1
circular-ribbon flare. It can be adequately explained by the LRC
model. The time lag between H$\alpha$ and SXR light curves suggests
that the driven energy originates from photospheric convection in
the low atmosphere before propagating to the middle (H$\alpha$) and
high (SXR) atmospheres.

\begin{acknowledgements}
We acknowledge the anonymous referee for their inspiring and
valuable comments. We thank the teams of NVST, GOES and SDO for
their open data use policy. This study is supported by NSFC under
grant 11973092, 11803008, 11790302, 11729301, the Youth Fund of
Jiangsu No. BK20171108, as well as National Natural Science
Foundation of China (U1731241), the Strategic Priority Research
Program on Space Science, CAS, Grant No. XDA15052200 and
XDA15320301. The Laboratory No. 2010DP173032. D.L. is also supported
by the Specialized Research Fund for State Key Laboratories and the
CAS Key Laboratory of Solar Activity (KLSA202003). S. Feng is
supported by the Joint Fund of NSFC (U1931107) and the key Applied
Basic Research program of the Yuanan Province (2018FA035).
\end{acknowledgements}

\end{document}